\documentclass{iopart}

\usepackage{amssymb}
\usepackage{graphicx}	% Include figure files
\usepackage{dcolumn}	% Align table columns on decimal point
\usepackage{bm}				% bold math

\begin{document}

%\begin{frontmatter}

% Title, authors and addresses

% use the thanksref command within \title, \author or \address for footnotes;
% use the corauthref command within \author for corresponding author footnotes;
% use the ead command for the email address,
% and the form \ead[url] for the home page:
% \title{Title\thanksref{label1}}
% \thanks[label1]{}
% \author{Name\corauthref{cor1}\thanksref{label2}}
% \ead{email address}
% \ead[url]{home page}
% \thanks[label2]{}
% \corauth[cor1]{}
% \address{Address\thanksref{label3}}
% \thanks[label3]{}

\title{An alternative method for simulating particle suspensions using lattice Boltzmann}

% use optional labels to link authors explicitly to addresses:
% \author[label1,label2]{}
% \address[label1]{}
% \address[label2]{}

\author{Lu\'is Orlando Emerich dos Santos}
\address{Center of Mobility Engineering, Federal University of Santa Catarina, 88040-900, Florian\'opolis, Santa Catarina, Brazil}
\ead{emerich@lmpt.ufsc.br}

\author{Paulo Cesar Philippi}
\address{Mechanical Engineering Department, Federal University of Santa Catarina, 88040-900, Florian\'opolis, Santa Catarina, Brazil}
\ead{philippi@lmpt.ufsc.br}

\begin{abstract}
In this study, we propose an alternative way to simulate particle suspensions using the lattice Boltzmann method. The
main idea is to impose the non-slip boundary condition in the lattice sites located on the particle boundaries. The focus
on the lattice sites, instead of the links between them, as done in the more used methods, represents a great simplification in
the algorithm. A fully description of the method will be presented, in addition to simulations comparing the proposed method
with other methods and, also, with experimental results.
\end{abstract}

%Uncomment for PACS numbers title message
%\pacs{00.00, 20.00, 42.10}
% Keywords required only for MST, PB, PMB, PM, JOA, JOB? 
\vspace{2pc}
\noindent{\it Keywords}: Lattice Boltzmann method, Particulate flow, Particle-fluid interactions
% Uncomment for Submitted to journal title message
%\submitto{\JPA}
% Comment out if separate title page not required
% main text

\section{Introduction} \label{Intro}

Particulate flows are found in many industrial processes 
\cite{Behrend1995,Glowinski1999,Patankar2000,Ding2003},
and have been subjected to considerable scientific investigation.
Recently, computer simulations have become an effective tool in these studies and
several methods have been applied, such as finite element method \cite{Feng1994},
Lagrange-multipliers \cite{Glowinski1999,Patankar2000}, direct forcing
method \cite{Yu2007}, lattice Boltzmann methods 
\cite{Behrend1995,Ding2003,Ladd1994,Ladd1994a,Aidun1995,Aidun1998,Ladd2001,Chopard2002,Nguyen2002,Feng2004,NGUYEN2005,Kromkamp2006},
solving the Stokes equation near
the particles (Physalis) \cite{Zhang2005} and combining two or more methods
\cite{Feng2004,Feng2005,Wang2008}.\\
In this study, we propose an alternative way to simulate suspensions
using the lattice Boltzmann method. The main idea is to impose the 
non-slip boundary condition in the lattice sites representing the
particle boundaries. The focus on lattice sites, instead of the link between
them, as done in the more used methods \cite{Ladd2001}, represents
a great simplification in the algorithm. Similar approaches, focusing
on the lattice sites were already tried \cite{Behrend1995}, although without popularity 
due, probably, to difficulties several in the implementation.\\
Its important to mention that the simplifications we propose
can reduce the accuracy in describing the details of the flow near
the particles, and these details can or cannot be important, depending
on the problem one wants to simulate. 
In the simulations we have done until now, in despite of the 
simplifications, the results we obtain are similar to the results
obtained by other methods (see section \ref{Validation}).

\section{The model} \label {model}

In this section we introduce the lattice Boltzmann method and 
present the model describing the particle-fluid
interaction. Interactions among particles
and between a particle and a solid surface will also be
discussed in this section.

\subsection{The lattice Boltzmann model} \label{LBM}

\begin{figure}
	\centering
		\includegraphics[width=0.45\textwidth]{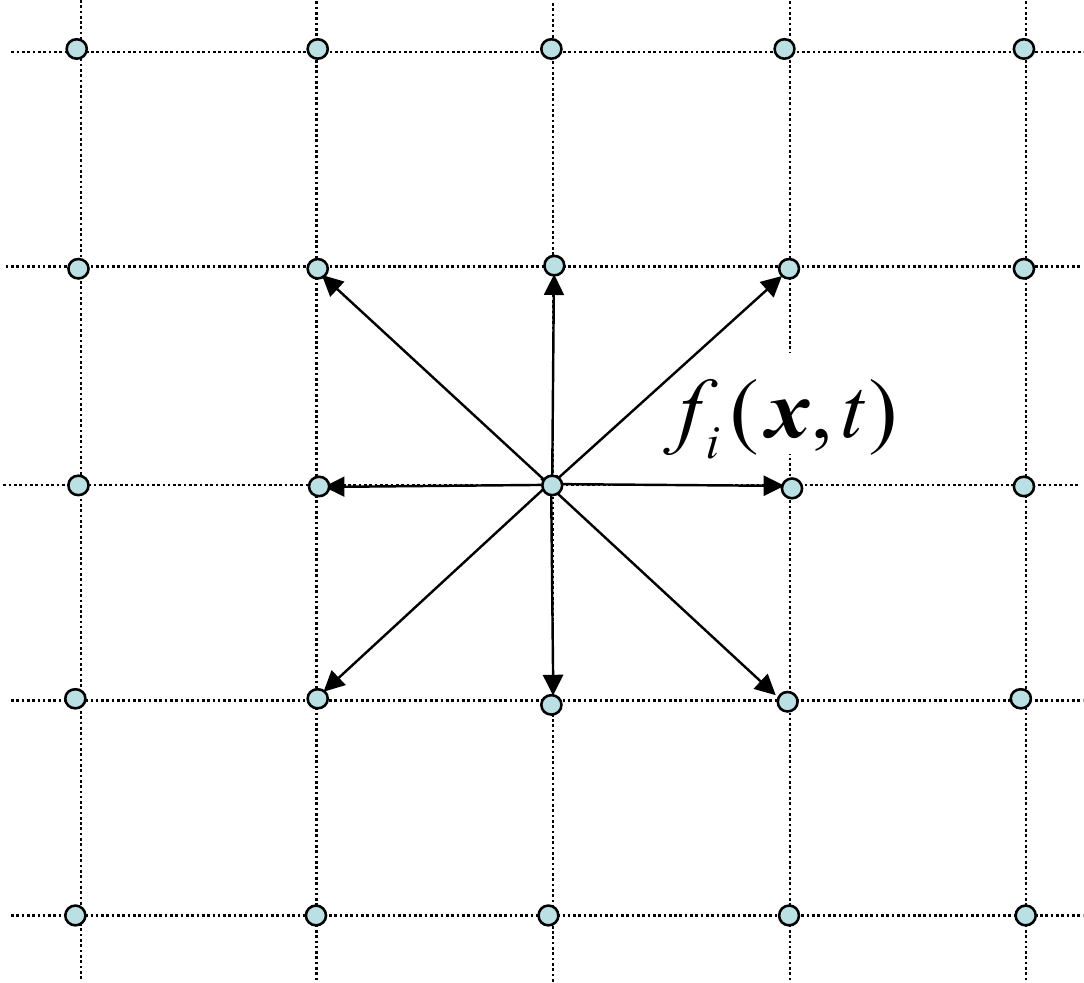}
		\caption{The D2Q9 lattice, used in two-dimensional simulations. The arrows 
		represent the distributions $f_i(\bm{x},t)$, and circles represent the lattice sites.}
	\label{fig:fig02}
\end{figure}

The lattice-Boltzmann method (LBM) is based on the discretization of the 
Boltzmann's mesoscopic equation\cite{He1997,Shan2006,Philippi2006}, 
usually with the BGK approach for the 
collision operator (for a comprehensive review see \cite{Succi2001}). 
In the LBM scale, the system is described  using 
a single particle distribution function, $f_i (\bm {x},t)$, representing 
the number of particles with velocity $\bm {c}_i $ at the site $\bm {x}$ 
and time $t$, where $i=0,...,b$. The particles are restricted to a 
discrete lattice, in a manner that each group of particles can move only in 
a finite number $b$ of directions and with a limited number of velocities 
(see Fig. \ref{fig:fig02}). 
Therefore, physical and velocity space are discretized. 
The local macroscopic properties such as total mass (the particle mass, m, 
is assumed unitary), $\rho (\bm {x},t)$, and total momentum, $\rho (\bm 
{x},t)\bm {u}(\bm {x},t)$, can be obtained from the distribution function 
in the following way:
\begin{equation}
	\rho (\bm {x},t)=\sum\limits_i {f_i } (\bm {x},t),
\end{equation}

\begin{equation}
	\rho (\bm {x},t)\bm {u}(\bm {x},t)=\sum\limits_i {f_i } (\bm {x},t)\bm {c}_i .
\end{equation}
The Lattice Boltzmann equation, that is, the discrete version of the 
Boltzmann equation with the BGK collision, operator is written 

\begin{equation}
	f_i (\bm {x}+ \delta_t \bm {c}_i,t+\delta_t)-f_i (\bm {x},t)=-\frac{\delta_t}
	{\tau }\left[ {f_i (\bm {x},t)-f_i^{eq} (\rho ,\bm {u})} \right]
\end{equation}
where, $\delta_t$ is the time step and $f_i^{eq} (\rho ,\bm {u})$ is a 
polynomial approximation of the Maxwell-Boltzmann equilibrium 
distribution \cite{Abe1997,He1997,Shan2006,Philippi2006}, 
a function of the local 
variables $\rho (\bm {x},t)$ and 
$\bm {u}(\bm {x},t)$. It can be shown, through a Chapman-Enskog analysis, 
that this system macroscopically will evolve according to the Navier-Stokes 
equations with a kinematic viscosity given by
\begin{equation}
\label{eq2}
\nu =c_s^2 \left( {\tau -1/2} \right),
\end{equation}
where $c_s $ is the sound velocity, a constant depending on the set of 
velocities $\bm {c}_i $.

\subsection{Particle-fluid interaction} \label{part_fluid}

\begin{figure}
	\centering
		\includegraphics[width=0.45\textwidth]{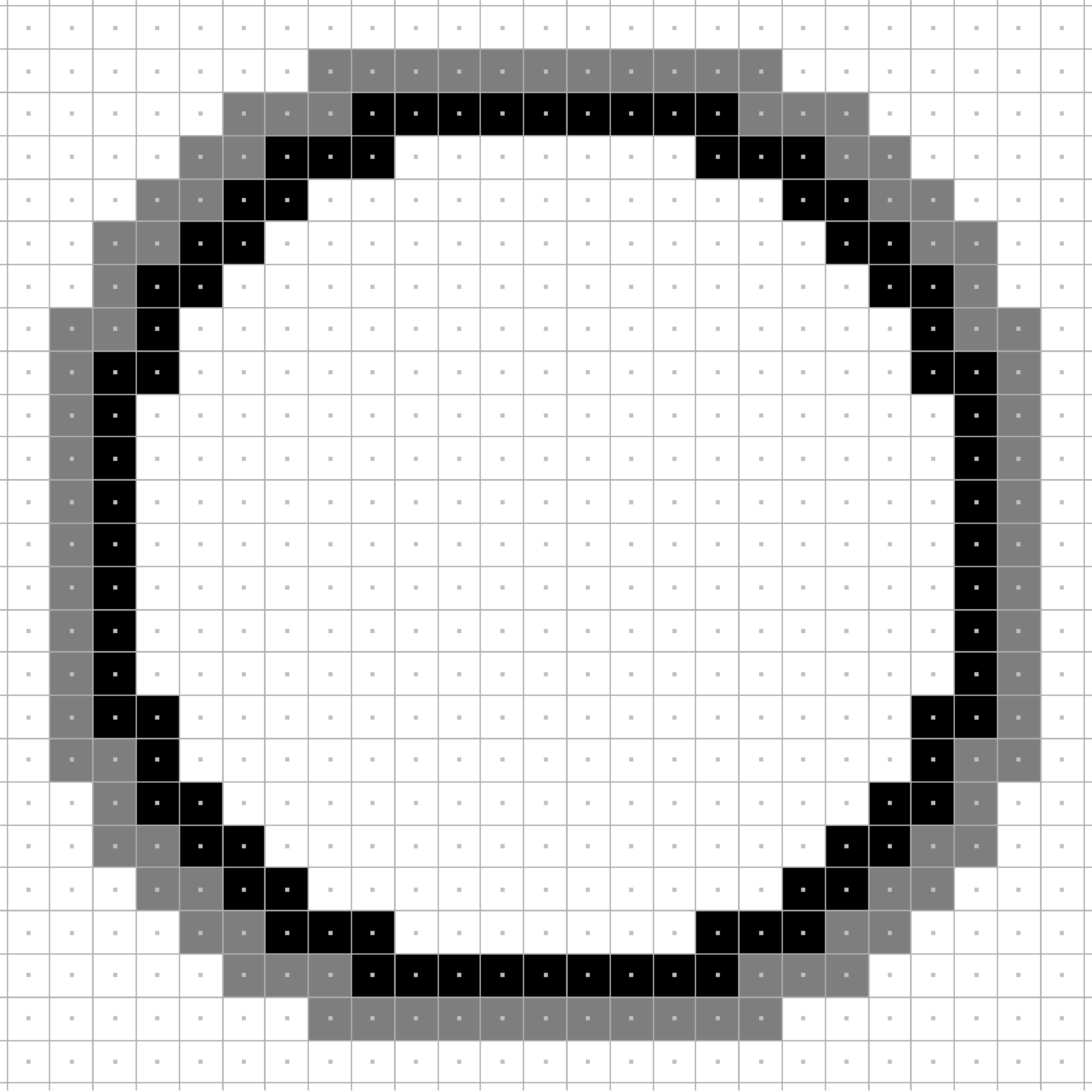}
	\caption{Boundary sites characterizing a particle. 
	The boundary sites (BS) are depicted in dark gray and the internal sites (IS), in dark. 
	The particle's boundary are regard to be between the IS and BS sites.}
	\label{esfera}
\end{figure}

The basic idea of the method  is that the fluid
in contact with a solid surface must acquire the velocity of this
surface, considering the non-slip condition. Keeping this in mind, a set
of boundary sites can be used to describe the particle. This approach is
similar to the one presented in \cite{Ladd1994} and \cite{Ladd2001}, although we focus in the lattice 
sites and not in the links between them. We denote \textit{boundary sites} (BS)
the particle sites in contact with fluid, and \textit{internal sites} (IS)
the particle sites in contact with the boundary sites (see Fig.\ref{esfera}). 
The particle's boundary is regarded to be halfway between the IS and the BS sites.
A particle will be represented by its central position $\bm{x}_p$, a radius $r_p$, 
a mass $m_p$ and a moment of inertia $I_p$. Its velocity will be denoted $\bm{u}_p$, 
and the angular velocity $\bm{\omega}_p$. Only spherical particles will be considered.

The presence of a particle will be represented only by its effects on the 
fluid in contact with the particle, altering the dynamics
of the sites BS and IS. 
The position of these sites will be denoted by $\bm{x}_B$ and $\bm{x}_I$,
respectively, and its integer counterpart (or the nearest integer of
each component of the vectors)
by $\left[\bm{x}_B\right]$ and $\left[\bm{x}_I\right]$, respectively.
Velocities and densities in the sites $\left[\bm{x}_B\right]$ and $\left[\bm{x}_I\right]$
will be denoted $\bm{u}_B$, $\bm{u}_I$ and $\rho_B$, $\rho_I$.
We emphasize that all sites will be updated by the usual collision/propagation steps,
but the IS and BS sites, in addition to these steps, will be
submitted to a set of substeps included between the collision and the propagation steps.
These substeps are presented in the sequence.

\begin{itemize}
	\item \textsl{Particle-fluid momentum transfer}
\end{itemize}

This substep is executed only in the BS sites.
As was previously mentioned, the velocity in the BS sites must acquire the 
particle's velocity. This velocity, considering the angular velocity, is 

\begin{equation}
\bm{u}_B' =\bm{u}_p + (\bm{x}_B-\bm{x}_p) \times \bm{\omega}_p .
\end{equation}

To impose this velocity in the fluid, that is in lattice sites,
the equilibrium distribution is employed, 
setting 
$f_i \left(\left[\bm{x}_B\right],t\right) = f^{eq}_i(\rho, \bm{u}_B')$.

Naming $\bm{p}_{B}=\rho_{B}\bm{u}_{B}$ 
the linear momentum of a BS site,  
we compute
\begin{equation}
\Delta\bm{p}_B = \rho_B(\bm{u}_B'-\bm{u}_B),
\end{equation}
and 
\begin{equation}
\Delta\bm{l}_B = \rho_B(\bm{x}_B-\bm{x}_p) \times (\bm{u}_B'-\bm{u}_B),
\end{equation}
\noindent where $\Delta\bm{l}_B$ represents the change in the angular momentum
caused by the change in the linear momentum $\Delta\bm{p}_B$.

Finishing this step we have the total 
momentum exchanged, $\Delta\bm{P}_B$ and $\Delta\bm{L}_B$:
\begin{equation}
	\Delta\bm{P}_B =\sum_{BS}{\Delta\bm{p}_B} 
\end{equation}
\begin{equation}
	\Delta\bm{L}_B =\sum_{BS}{\Delta\bm{l}_B} 
\end{equation}

\begin{itemize}
	\item \textsl{Particle's acceleration}
\end{itemize}

This substep accounts for the change in particles velocity caused
by fluid. According to the Newton's third law of motion, we simply
compute

\begin{equation}
	\bm{u}_p' = \bm{u}_p - \Delta\bm{P}_B / m_p, 
\end{equation}

\begin{equation}
	\bm{\omega}_p' = \bm{\omega}_p - \Delta\bm{L}_B / I_p,
\end{equation}

\begin{equation}
\bm{u}'_I =\bm{u}_p' + (\bm{x}_I-\bm{x}_p) \times \bm{\omega}_p' .
\end{equation}

\begin{itemize}
	\item \textsl{Updating of particle's position}
\end{itemize}

Due to the spherical symmetry it is not necessary
to take into account the rotation of the particles. The position
is updated by doing:

\begin{equation}
	\bm{x}_p' = \bm{x}_p + \delta_t \bm{u}_p'.
\end{equation}

The positions of boundary and internal sites, also, must be updated:

\begin{equation}
	\bm{x}_B' = \bm{x}_B + \delta_t \bm{u}_p';
\end{equation}

\begin{equation}
	\bm{x}_I' = \bm{x}_I + \delta_t \bm{u}_p'.
\end{equation}

Clearly, this procedure includes errors of order ($\delta_t^2$), more
precise procedures could be employed. Nevertheless, there is 
a lot of imprecision in describing the shape of the particle, therefore
it is not necessary to be so precise in updating particle's position.

\begin{itemize}
	\item \textsl{Velocity of the internal sites}
\end{itemize}

To impose this velocity  the equilibrium distribution is employed again,
setting 
$f_i\left(\left[\bm{x}_I '\right],t\right) = f^{eq}_i(\rho, \bm{u}'_I)$.
\\
\\
Some comments are necessary to clarity the physics behind these sub-steps. 
Consider a particule that is found with a velocity $\bm{u}_p$ different of the fluid 
velocity around it, in the beginning of the particle-fluid interaction step. 
This situation never occurs in the continuum but this is possible in discrete models 
due to the discretization of time. Particle and fluid must then exchange linear and 
angular momentum until the fluid around the particle acquires the velocity of 
the particle surface. In this process the particle velocity also changes due 
to the fluid reaction on the particle and in accordance with Newton's third law 
of motion. Action and reaction happen simultaneously in the physical process, 
but in the discrete case this occurs in steps \textit{1} and \textit{2}. In the 
first step the particle transfers linear and angular momentum to the fluid. 
In consequence the fluid in contact with the particle surface particle accelerates 
until have acquired the same particle velocity. As it was earlier mentioned two 
hypotheses are used for this step: non-slipping on the particle surface 
and the hypotheses that the fluid-particle equilibration takes a time interval 
smaller than the time step used in the simulation. 
This enables the use of an equilibrium distribution to impose the velocities. 
With the fluid velocity altered, we then calculate the change in the linear and 
angular momentum of the fluid due to this acceleration. These variations must 
be the same as the corresponding changes for the particle, in accordance with 
the Newton's third law. We then proceed to step \textit{2} and recalculate the 
particle velocity. The velocity to be imposed in sites IS is also calculated in 
accordance with the new particle velocity.

In a third step, the position of the particle and sites IS and BS are changed. 
The spherical symmetry of the particles eases this step since only the translation 
velocities are taken into account in the calculation of the new positions. 
It must be stressed that although sites BS belong to the fluid phase, 
they have their position changed in accordance with the particle velocity. 
Indeed, these sites are in the particle boundary and must follow its displacement. 
In other words the displacement of the boundary sites is independent of the fluid 
particles that are occupying these sites, in a given instant.

Finally, in a fourth step, we impose the velocity $\bm{u}'_I$ to the sites IS. 
This step is important because during the propagation step the information 
in these sites is transferred to the adjacent sites BS. Therefore, the sites 
BS will have their velocity composed by the particle and fluid velocities and will define 
a new change of momentum in the following time step.

\subsection{Interaction between particles} \label{part_part}

\begin{figure}
	\centering
		\includegraphics[width=0.90\textwidth]{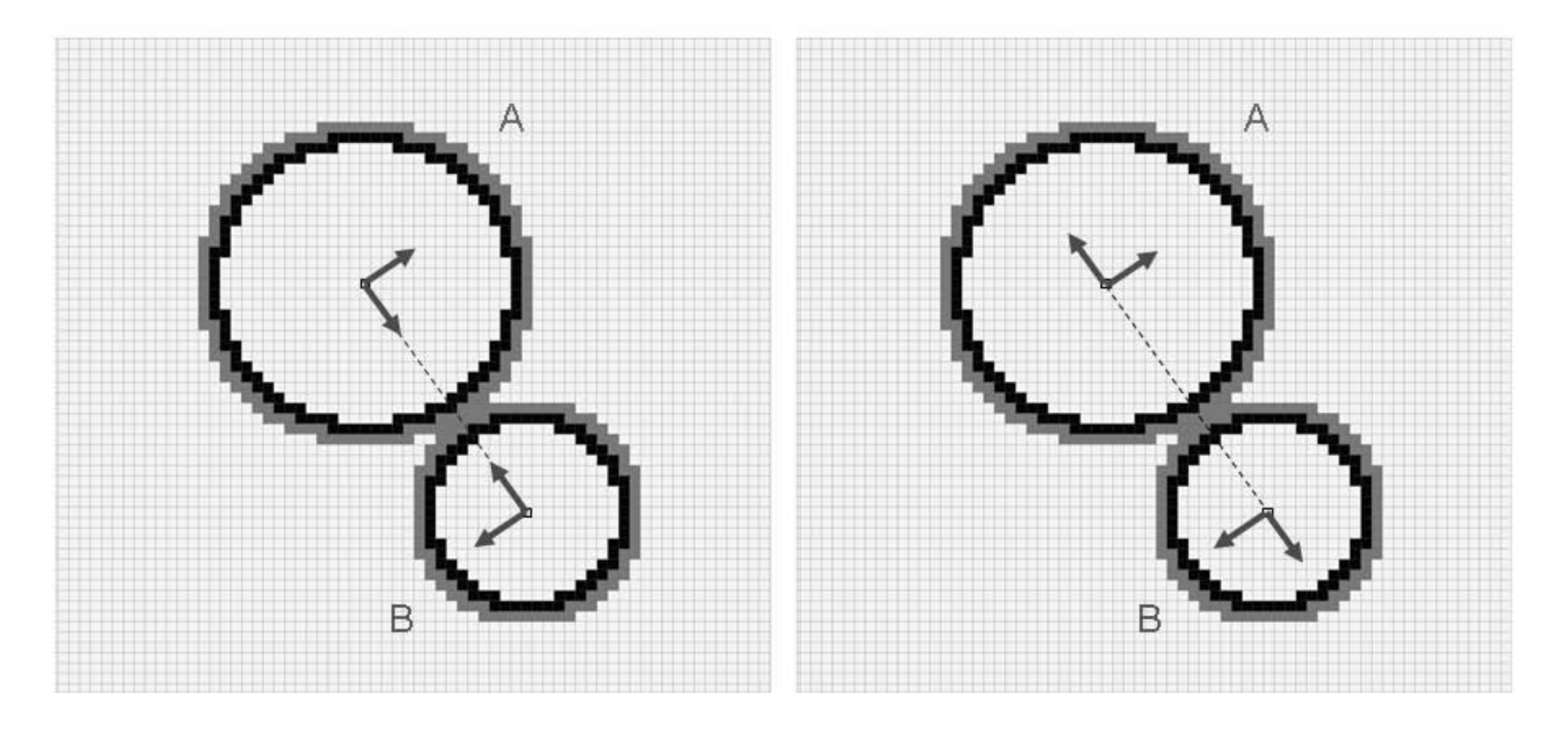}
	\caption{Collision between two particles, A and B. The arrows indicate the normal and tangential 
	velocities before and after the collision. The velocities are represented in the center of mass reference frame.}
	\label{Collision}
\end{figure}

Especially in simulations involving a great number of particles,
as the simulation presented in subsection \ref{504particles}, the interaction
between particles must carefully be addressed. There are many
techniques to treat these interactions, possibly the most popular
approach is to introduce a
repulsive force between particles when the gap  
between becomes smaller than a given threshold\cite{Feng2004}. We chose another approach
in this work, treating the collisions as occurring between completely
rigid particles (hard-core collisions). In this way it is not necessary to set
the parameters of repulsive forces. We simply impose (see Fig.\ref{Collision})\\

\begin{equation}
\bm{v}'_{An} = -\bm{v}_{An},
\end{equation}
\begin{equation}
\bm{v}'_{Bn} = -\bm{v}_{Bn},
\end{equation}

\begin{equation}
\bm{v}'_{At} = \bm{v}_{At},
\end{equation}
\begin{equation}
\bm{v}'_{Bt} = \bm{v}_{Bt},
\end{equation}
where all velocities are represented in the center of mass reference frame.
This imposition must be done before updating particles' positions, that is, it can
be considered as part of the particle's acceleration substep.\\
It's necessary to underline that the way we chose to 
introduce the particle-particle interaction is not new, nor regarded as part of 
the method. Moreover, this choice was done considering solely 
the easeness in the implementation. In accordance with the Reynolds number, 
volume fraction of the particles and the involved force, more elaborated 
techniques, employing lubrication\cite{Ladd1994a} or spring forces\cite{Hoefler2000} 
may reveal to be necessary.

\section{Validation}\label{Validation}

Four cases are presented in order to validate the model. The first case simulated was
the flow around a massive two-dimensional particle. The particle's mass was chosen in such 
way that it doesn't move, therefore it's possible to compare the flow around
the two dimensional particle and the flow around a solid cylinder, 
simulated using bounce-back. The drag coefficient for a massive particle was also computed considering 
several Reynolds number
and compared with drag coefficients obtained by other methods.  
The second case simulated was the flow of a neutrally 
buoyant two-dimensional particle in a shear flow, the results are compared
with the results presented by Feng and Michaelides\cite{Feng2004}. A simulation of a sphere settling in a 
closed box was done in order to validate the model in a three-dimensional 
simulation. The results obtained was compared with the ones presented by ten Cate et al.\cite{Cate2002}. 
Finally, it was simulated the sedimentation of 504 two-dimensional particles in  
an enclosure. 

\subsection{Flow around a massive two dimensional particle}

\begin{figure}
	\centering
		\includegraphics[width=0.90\textwidth]{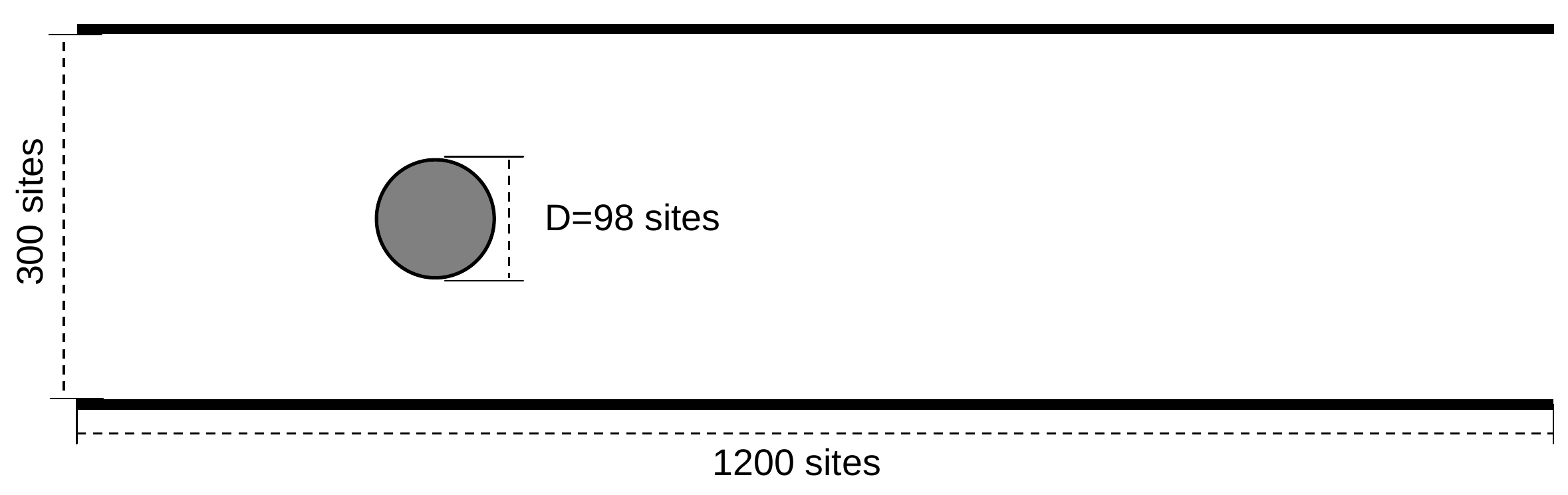}
	\caption{Geometry used to simulate the flow in a channel with an obstacle.}
	\label{geo_massive_ particle}
\end{figure}

\begin{figure}
	\centering
		\includegraphics[width=0.70\textwidth]{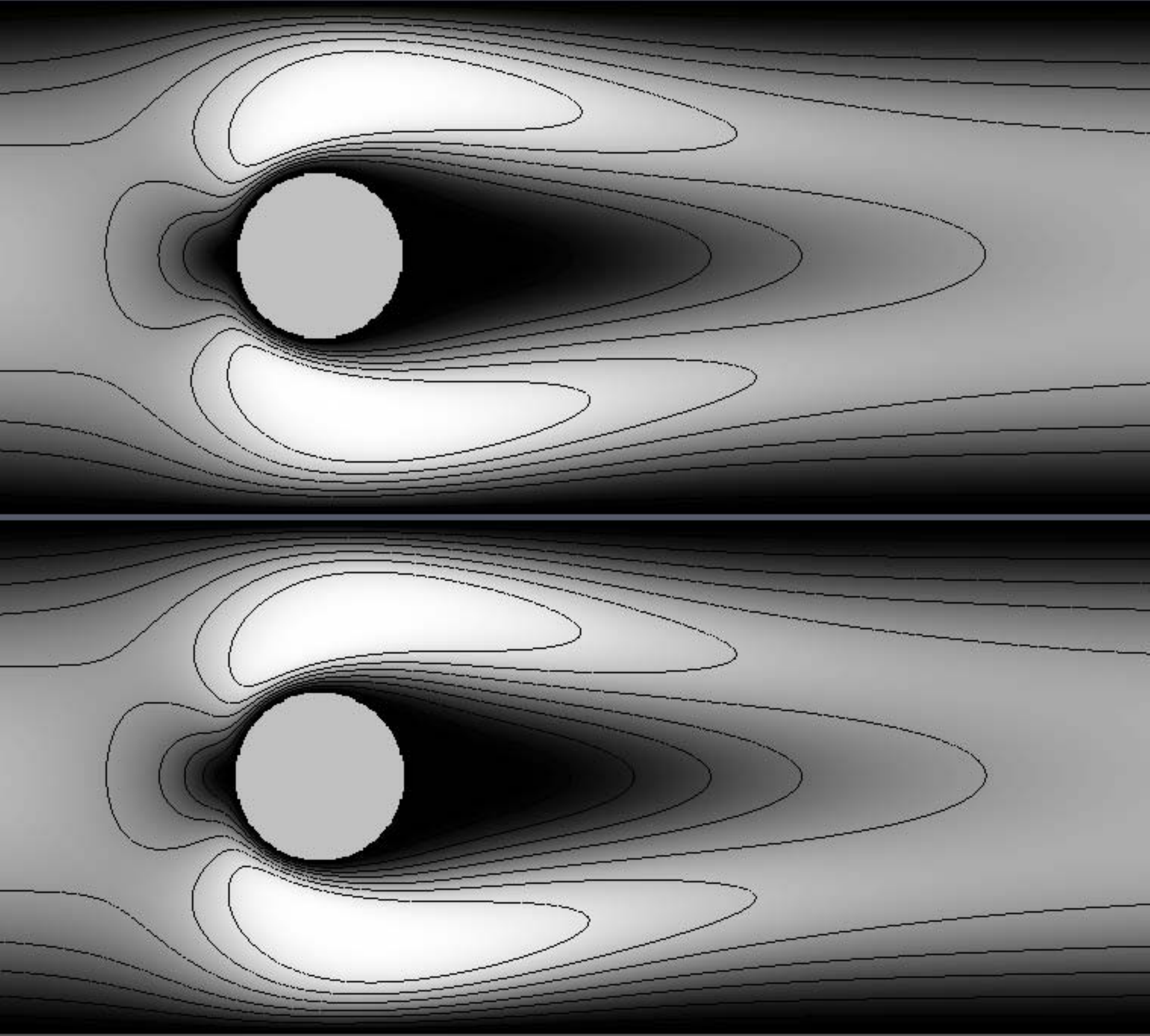}
	\caption{Top:flow around a solid cylinder. Bottom: flow around a massive particle. 
	The velocities vary from zero (black) to 0.075 (white)}.
	\label{massiveparticle2}
\end{figure}

\begin{figure}
	\centering
		\includegraphics[width=0.70\textwidth]{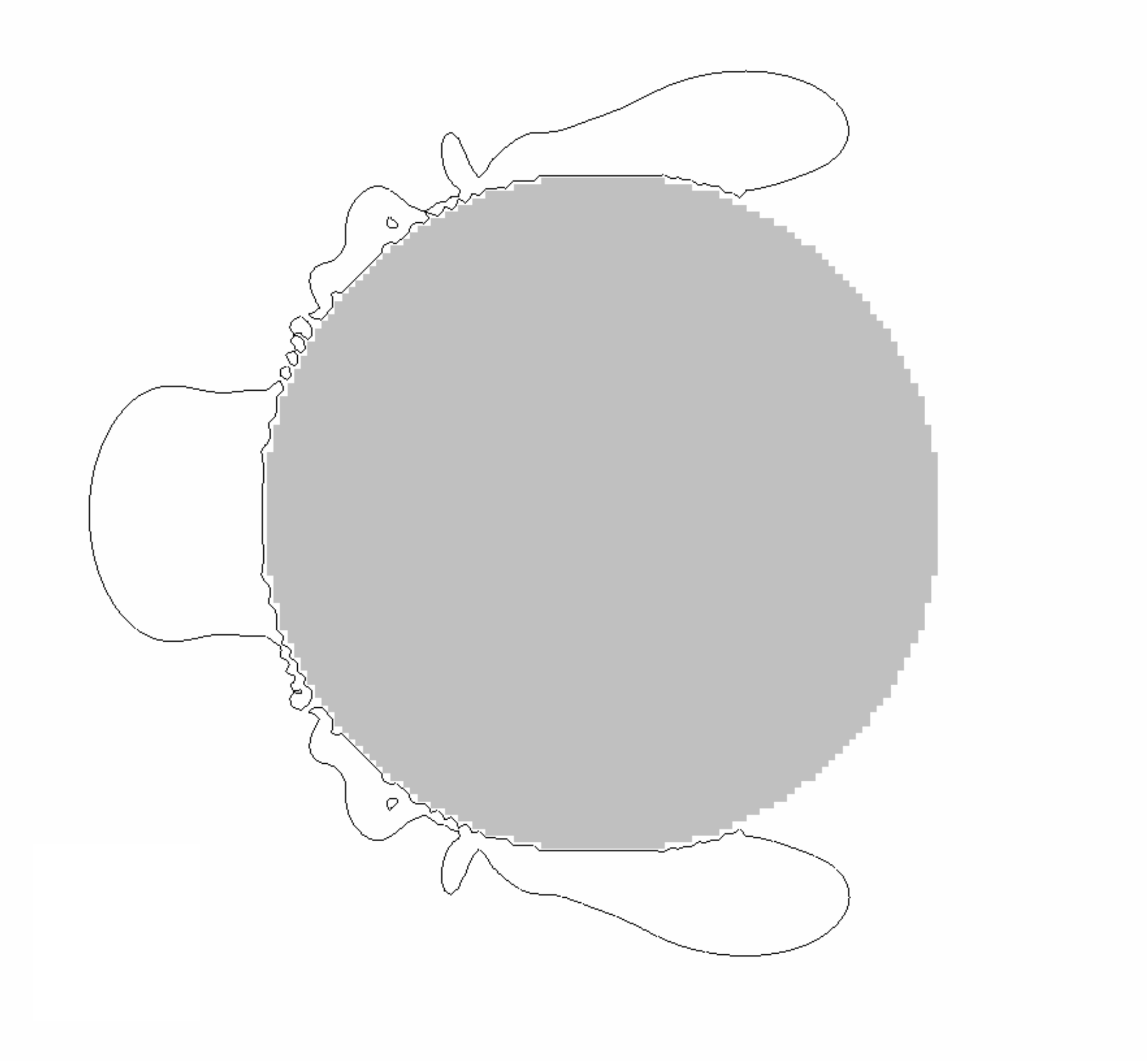}
	\caption{The difference in the velocities considering the 
	simulations using a solid cylinder and a massive particle. In the
	enclosed region the errors are of order $O(10^{-4})$, outside the errors are smaller than this.}
	\label{mass_part_diff}
\end{figure}

The first geometry used to simulate the flow around a massive particle is depicted in 
Fig. \ref{geo_massive_ particle}. The simulations of a flow around a solid 
cylinder was carried out using bounce-back boundary conditions and the simulation of
flow around a massive particle was done using the method proposed in this work, that
is, using the equilibrium distribution to impose the velocities in the boundaries
of the particle.
The results of both simulations are presented 
in Fig. \ref{massiveparticle2}. To emphasize the deviations 
we plot in Fig. \ref{mass_part_diff} the magnitude of the difference between the velocities.
In the enclosed regions of Fig. \ref{mass_part_diff} the deviations
are of order $10^{-4}$ (the velocities in the simulations varying from zero to $0.075$).
Is important to notice that, in despite of deviations that can appear near the
solid surface as result of applying different boundary conditions, 
the overall flow behavior is the same. Off course, depending on the applications
ones intends to focus, these differences must be taken into account.

\begin{figure}
	\centering
		\includegraphics[width=0.90\textwidth]{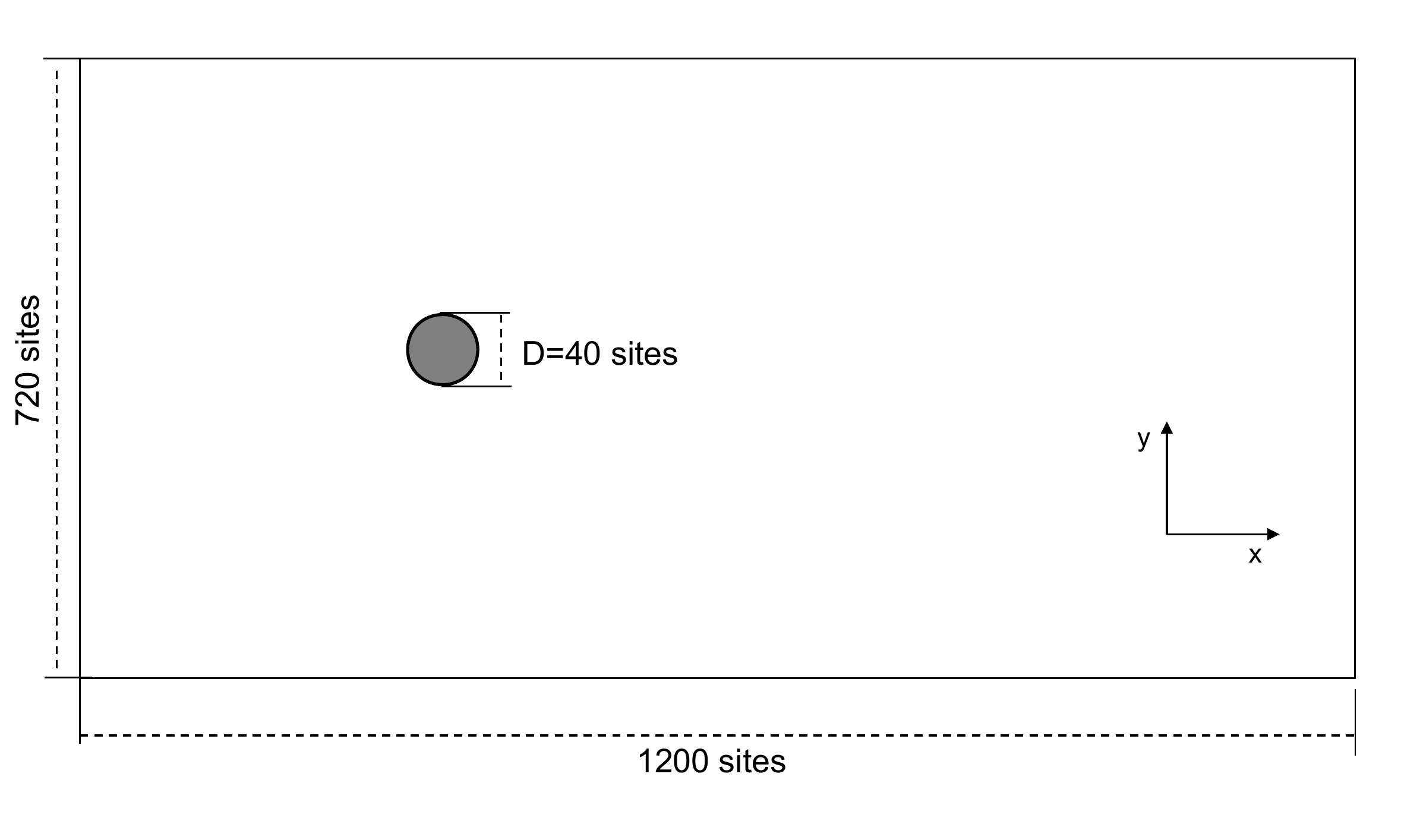}
	\caption{The geometry used to compute the drag coefficient of the 2D particle.}
	\label{geo_mass_part_drag}
\end{figure}

\begin{figure}
	\centering
		\includegraphics[width=0.85\textwidth]{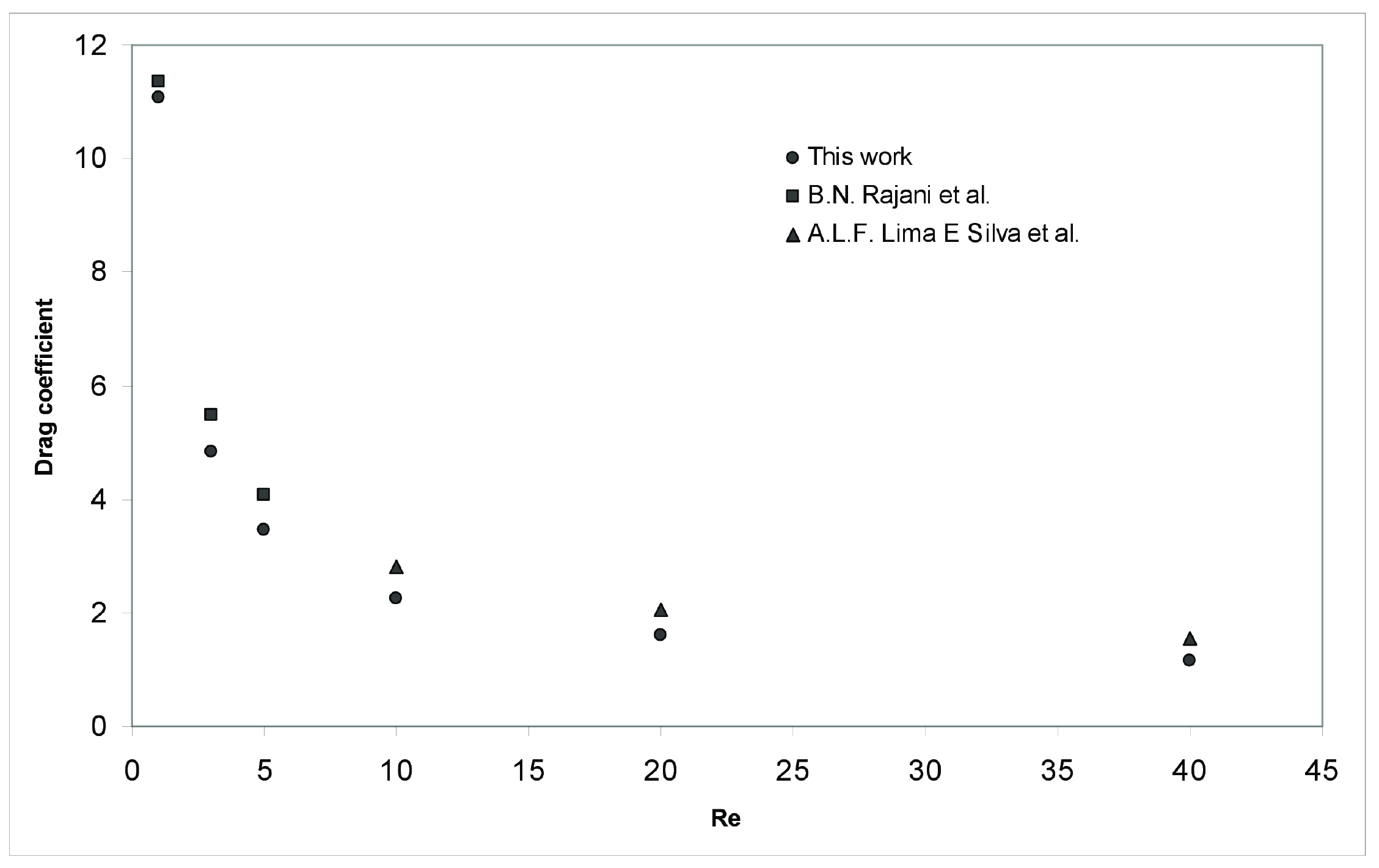}
	\caption{Drag coefficients for Reynolds number varying from one to forty. The
	results computed using the proposed model are compared with the work 
	by Rajani et al. \cite{Rajani2009} (Reynolds from one to five) 
	and Silva et al. \cite{Silva2003} (Reynolds from ten to forty).}
	\label{drag}
\end{figure}

Simulations to obtain drag coefficients were also carried out in the geometry presented
in Fig. \ref{geo_mass_part_drag}. Velocities varying from 0.00417 to 0.1667 in the x direction 
were imposed at the boundaries (left, right, top and bottom), resulting in the Reynolds number
varying from $Re=1$ to $Re=40$. The results obtained are shown in Fig. \ref{drag}, beside 
the results published by Rajani et al.\cite{Rajani2009} and Silva et al. \cite{Silva2003}.
Although the drag coefficients computed in this work were systematically lower than
the results obtained by other methods, the importance of these errors is dependent 
on what we want to describe in a given problem,as it can be seen in the simulations 
presented in the sequence.

\subsection{Neutrally buoyant two dimensional particle in a shear flow}

\begin{figure}
	\centering
		\includegraphics[width=0.95\textwidth]{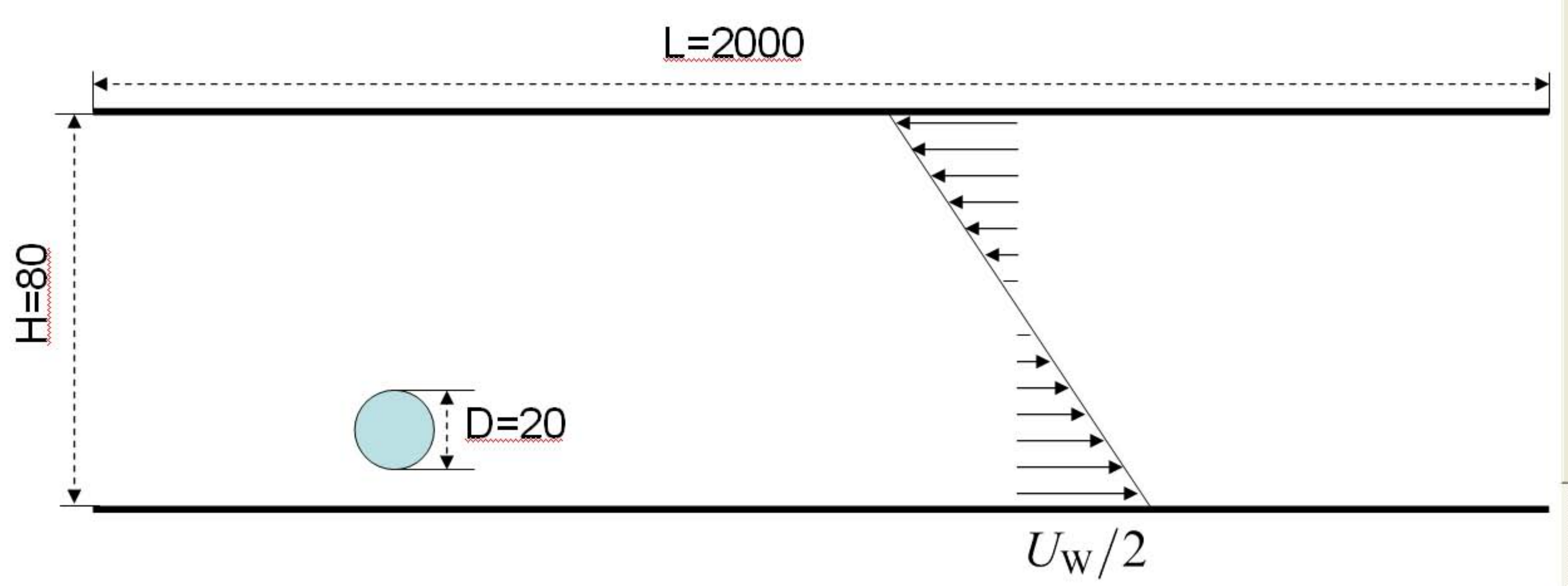}
	\caption{Schematic depiction of the problem simulated: a neutrally buoyant
	two dimensional particle in a shear flow}
	\label{geometria_couette}
\end{figure}

\begin{figure}
	\centering
		\includegraphics[width=0.85\textwidth]{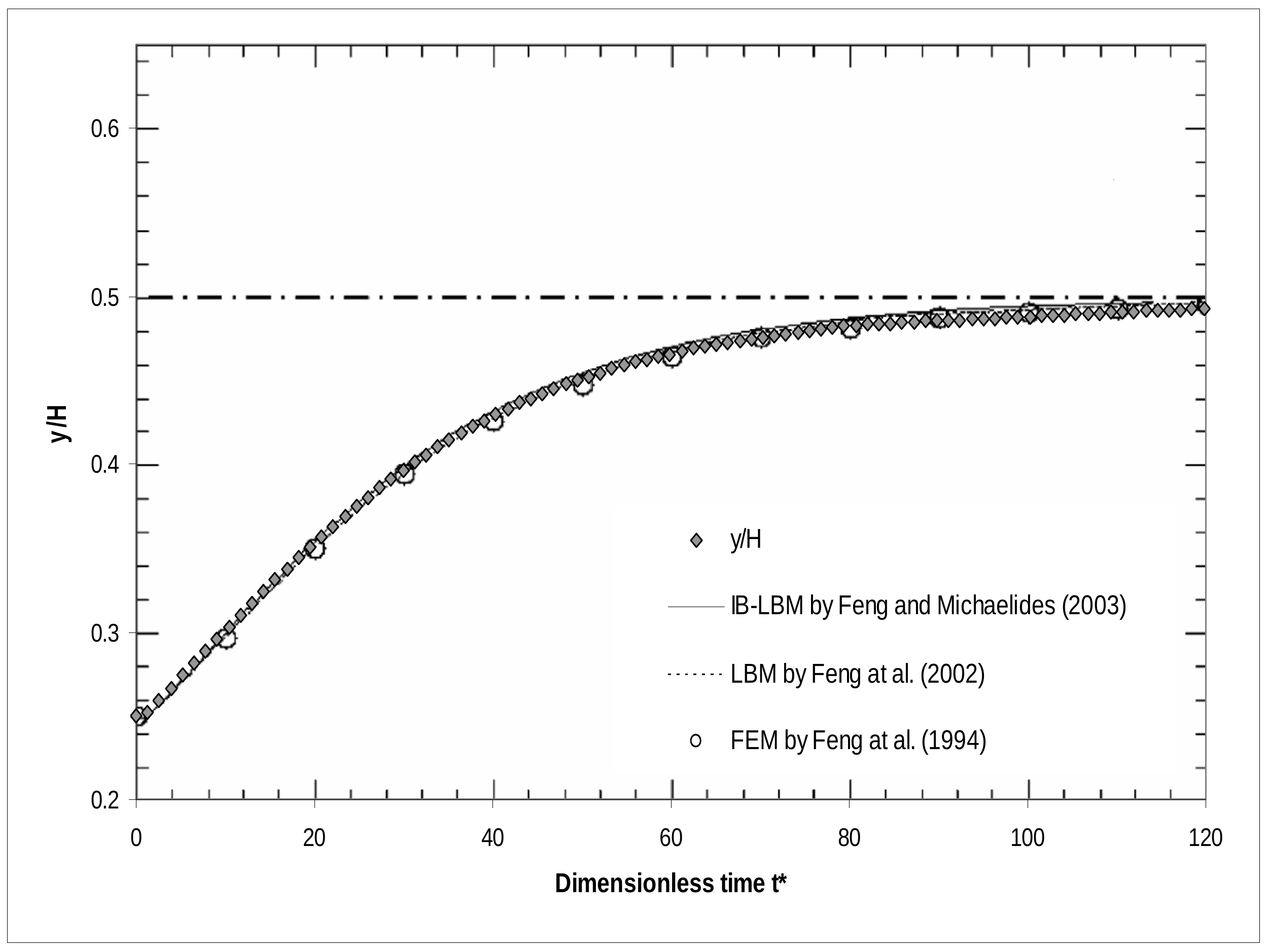}
	\caption{A neutrally buoyant particle in a shear flow --
	comparison of the results obtained using the proposed model for fluid-solid interaction and previous models.
	The y-axis shows the particle's position with respect to the lateral of channel divided it's width H.}
	\label{couette_results}
\end{figure}

The motion of a neutrally buoyant two-dimensional particle moving in viscous 
fluid was already simulated using LBM \cite{Feng2002},\cite{Feng2002a},\cite{Feng2004} 
as well 
as using finite element method \cite{Feng1994} , and was chosen as one
of the validations of the present model. The geometry of the problem 
is described in Fig. \ref{geometria_couette}, where $U_w/2$ and $-U_w/2$ 
are the velocities imposed. 
Periodic boundary conditions are imposed in the left and right boundaries. 
The relaxation time was set $\tau=0.6$, which implies a kinematic viscosity $\nu=1/30$, 
in lattice units.
The parameters chosen are the same used in Feng \& Michaelides paper\cite{Feng2004}, 
in order to ease the comparison. 
The velocity equals $U_w/2=1/120$, therefore, the shear rate for the flow is $\gamma = U_w/H = 1/120$, 
and a dimensionless time, $t^*$, is defined $t^*=t\gamma^2r_s/\nu$, where $r_s$ is the particle radius.
Fig. \ref{couette_results}, shows the migration of the particle, initially at the position $y=H/4$, 
toward the center. The agreement between the results using the proposed model and the previous models 
are quite good.

\subsection{A sphere settling in a closed box}

In this subsection the trajectory and velocity of a sphere
settling in a closed box is
simulated using the model here prosed and compared
with experimental results obtained by Cate et al. \cite{Cate2002}. 
\begin{figure}
	\centering
		\includegraphics[width=0.80\textwidth]{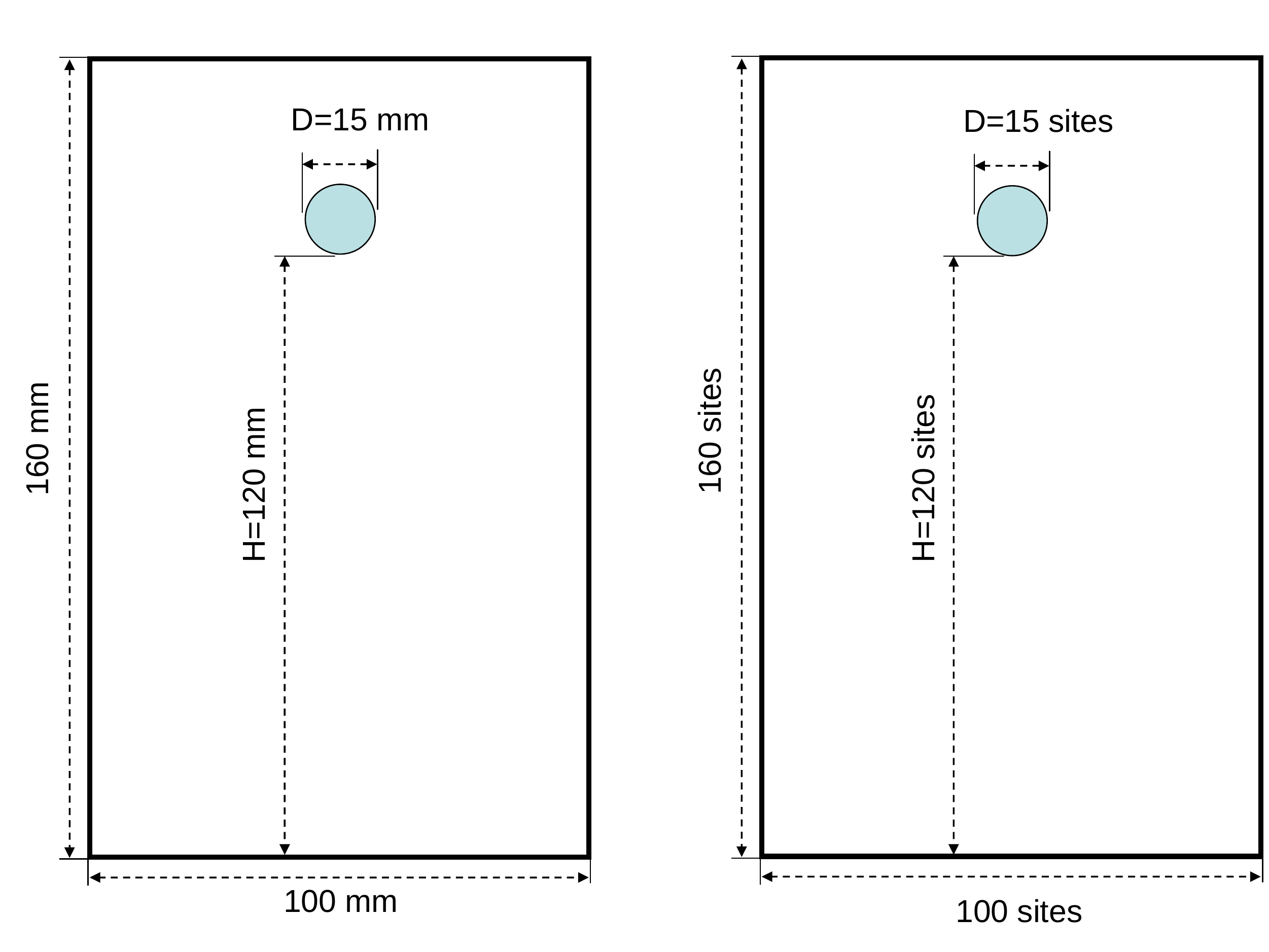}
	\caption{Schematic depiction of the geometry used in the experiments (left) and 
	in the simulations (right).}
	\label{geo_particle_set}
\end{figure}
The settling sphere has
a diameter of $D=15 mm$ and density $\rho = 1120 kg/m^3$.
The container dimensions are depth$\times$width$\times$height$= 100\times100\times160 mm$ 
(see Fig. \ref{geo_particle_set}). Four cases were simulated considering the
different densities and viscosities of the fluid in which the sphere will settle, 
corresponding to a Reynolds number varying from $Re=1.5$ to $Re=32.2$.
The fluid characteristics and the parameters used in the simulations are shown in Table \ref{tab1}.

\begin{table}[!h]
\begin{center}
\caption{Fluid properties in the experiment and parameters used in simulations.}

\begin{tabular}{ |c | c | c | c | c | }
\hline\hline
   			 &  $\; \rho _f[kg/m^3]\;$ & $\mu _f [10^{-3}Ns/m^2]$ & $\ \;\;\tau \;\;\ $  & $\;\delta_t [10^{-3}s]\;$\\
\hline
  Case E1		 &	970						& 		373							 & 	0.9  & 	0.347						 \\
\hline
  Case E2		 &	965						& 		212							 & 	0.9 & 	0.607						 \\
\hline
  Case E3		 &	962			 		  &	 		113						 	 & 	0.8 & 	0.851 					 \\
\hline
  Case E4		 &	960			 			& 		58						  & 	0.9 & 	2.207						 \\
\hline\hline
\end{tabular}\label{tab1}
\\
\end{center}
\end{table}

\begin{figure}
	\centering
		\includegraphics[width=0.90\textwidth]{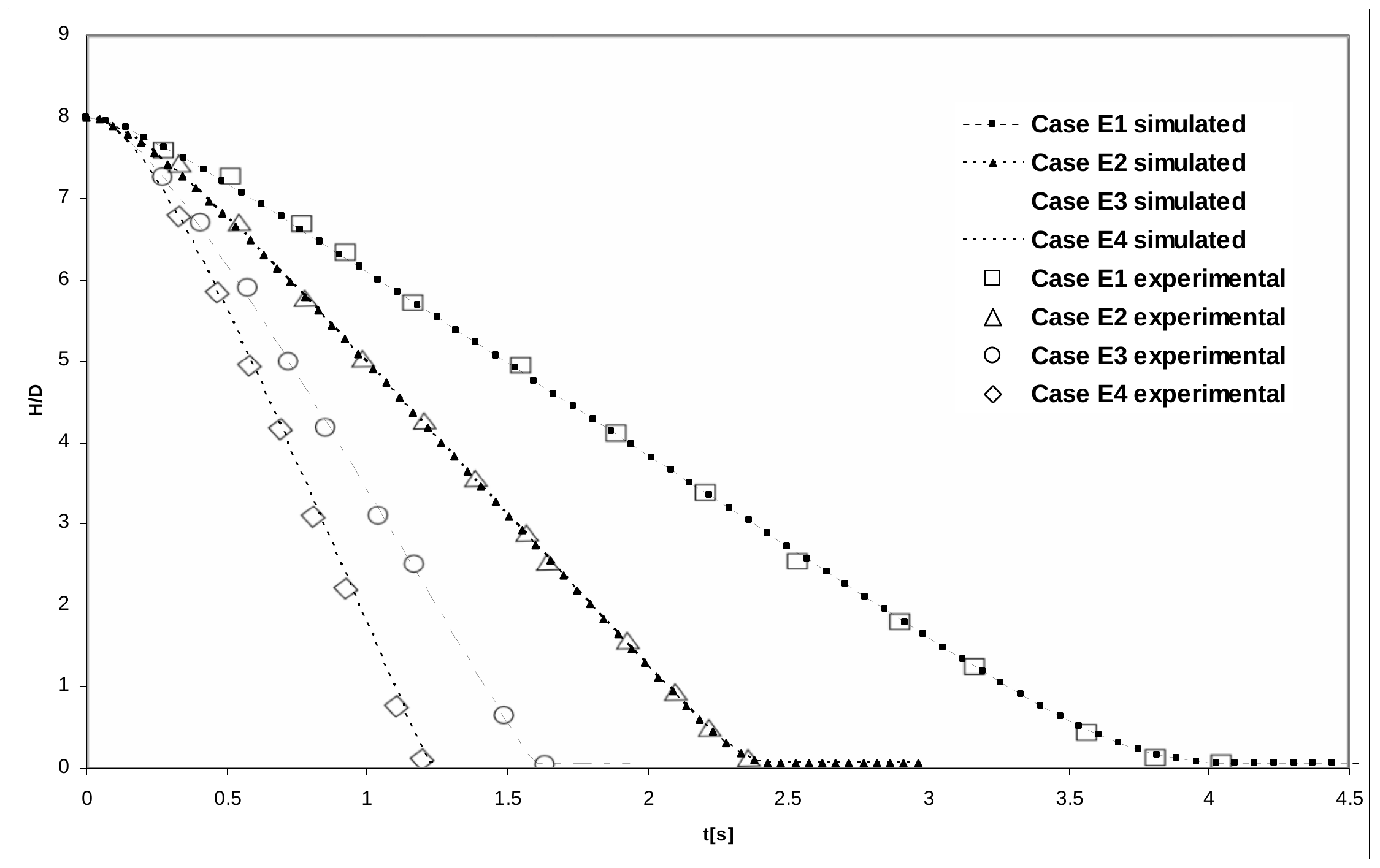}
	\caption{The position of the particle settling in a closed box. 
	The simulated results are compared with the experimental results from Cate el al. (2002).
	The y-axis presents particle's position H (see Fig. \ref{geo_particle_set}), divided by it's diameter D.
	The x-axis shows time in seconds.}
	\label{set_results}
\end{figure}

\begin{figure}
	\centering
		\includegraphics[width=0.90\textwidth]{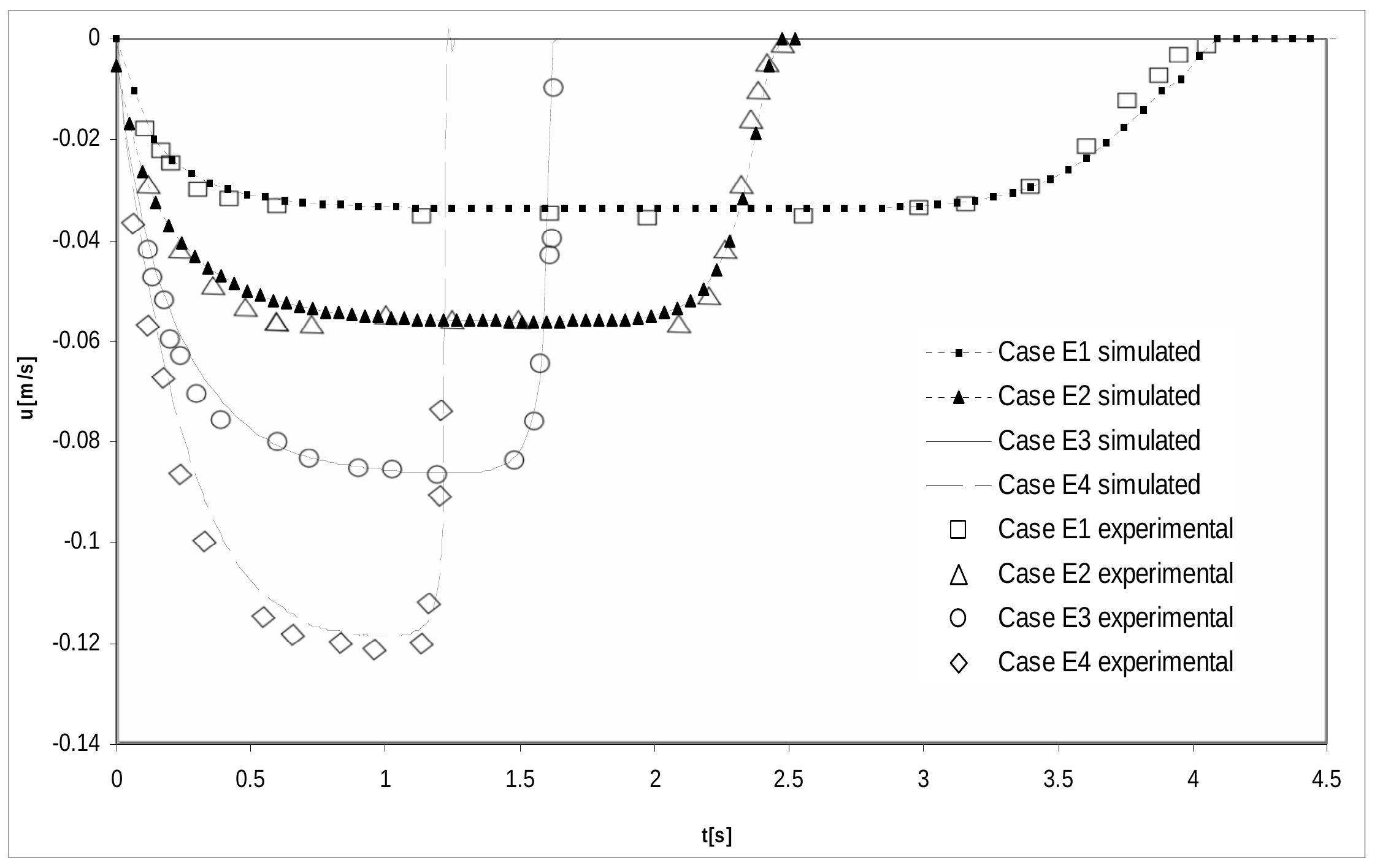}
	\caption{Comparison between simulation and the experimental results from Cate el al. (2002).
	The y-axis shows the velocity of the settling particle in meters per second and the x-axis
	indicates time in seconds.}
	\label{set_results2}
\end{figure}

Figure \ref{set_results} and \ref{set_results2} present comparisons between simulation and experiment
for the trajectories and velocities of the settling particle, respectively.

\subsection{504 particles settling in a closed box}\label{504particles}

\begin{figure}
	\centering
		\includegraphics[width=1.\textwidth]{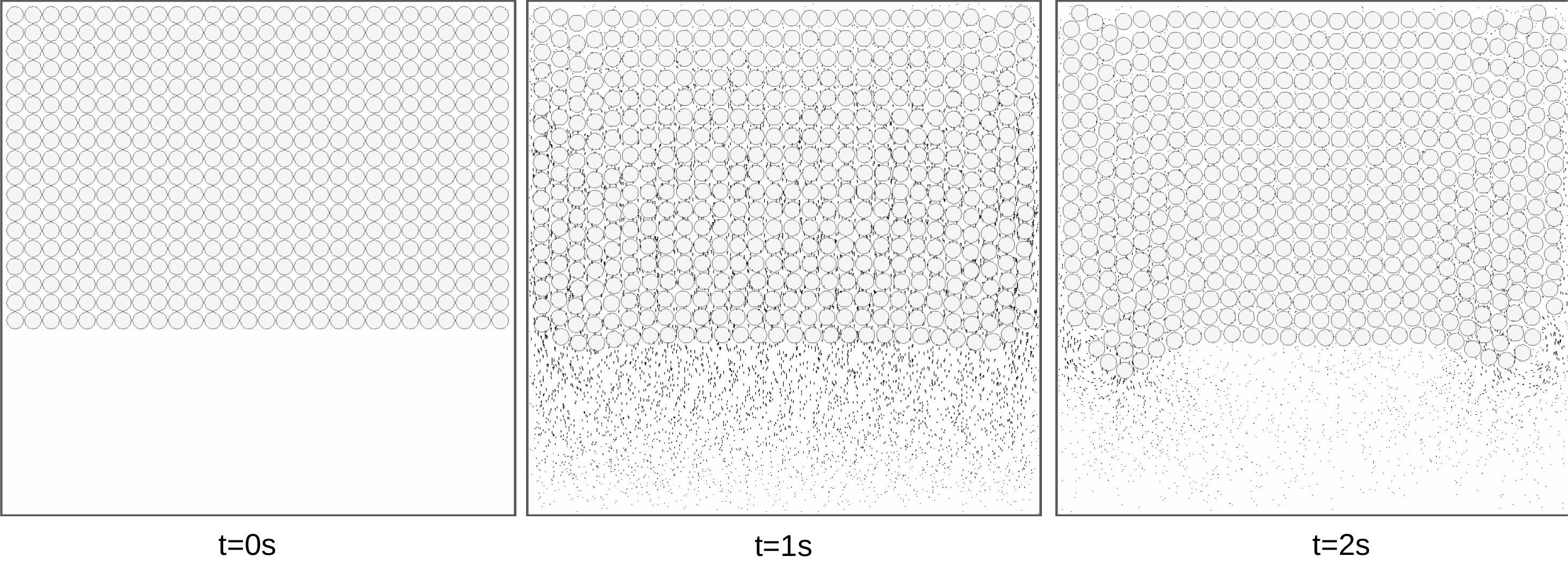}
	\caption{Configurations of the settling particles.}
	\label{504part_0_2}
\end{figure}

\begin{figure}
	\centering
		\includegraphics[width=1.\textwidth]{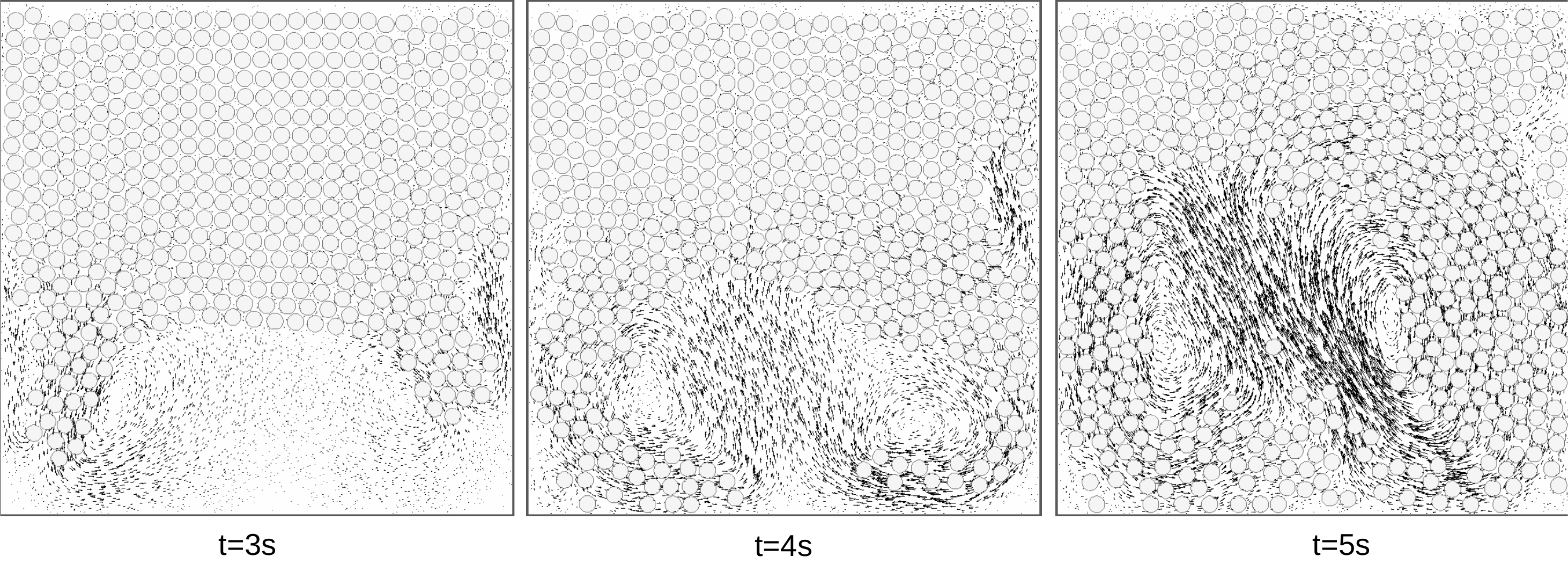}
	\caption{Configurations of the settling particles.}
	\label{504part_3_5}
\end{figure}

\begin{figure}
	\centering
		\includegraphics[width=1.\textwidth]{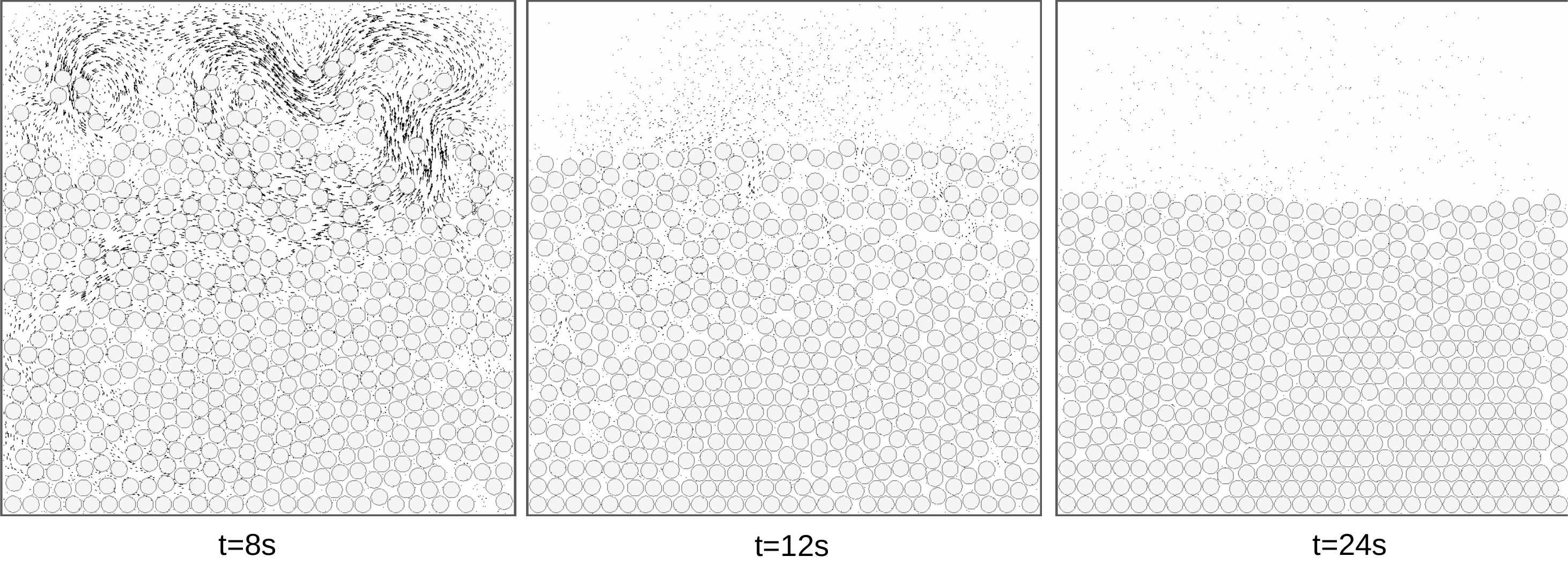}
	\caption{Configurations of the settling particles.}
	\label{504part_8_24}
\end{figure}

The problem of a large number of particles settling in a closed 2D box was already simulated
by other methods \cite{Glowinski1999}, \cite{Feng2004}. All the parameters were chosen in order
to compare with the previous works. That is, 504 circular particles with diameter $D = 0.625 cm$
settling in box having $10^{-2} m$ width and $10^{-2} m$ height. The fluid and particle densities are 
$\rho_f=1000kg/m^3$ and $\rho_p=1010kg/m^3$, respectively, and the fluid kinematic viscosity is $\nu_f=10^{-6} m^2/s$.
Representing the box by $512\times512$ sites and using a relaxation time $\tau=0.9915$ we will have
a time step of $0.00025 s$. The process of sedimentation simulated from the initial state
to $24s$ is presented in Fig. \ref{504part_0_2}, Fig. \ref{504part_3_5} and Fig. \ref{504part_8_24}.
The figures show the development of a Rayleigh--Taylor instability and are, qualitatively, in
agreement with the previous works. The differences between the three simulations 
(the one presented here and the simulations of refs. \cite{Glowinski1999} and
\cite{Feng2004}) are, possibly, a result of the differences in treating the 
collisions between particles and differences arising from compressibility effects
that are present in the lattice Boltzmann methods.

\section{Conclusion}

In this study an alternative and simpler way to simulate
particle-fluid interactions is proposed. The lattice Boltzmann
method is  employed to simulate the fluid flow and the 
particles are simulated using the Newton's law. The coupling
is made applying the equilibrium distribution function to 
assure the non-slip condition near the solid surfaces. 
Several simulations are presented showing that
the method can simulate particle-fluid interactions with a precision
comparable with other methods.

% The Appendices part is started with the command \appendix;
% appendix sections are then done as normal sections
% \appendix

% \section{}
% \label{}

%\begin{thebibliography}{00}

\section*{References}

\bibliography{Particles_new}
\bibliographystyle{spiebib}

% \bibitem{label}
% Text of bibliographic item

% notes:
% \bibitem{label} \note

% subbibitems:
% \begin{subbibitems}{label}
% \bibitem{label1}
% \bibitem{label2}
% If there is a note, it should come last:
% \bibitem{label3} \note
% \end{subbibitems}

%\bibitem{}

%\end{thebibliography}

\end{document}